\documentclass{aa}
\input{epsf}
\usepackage {graphicx}
\usepackage {longtable}

\begin{document}
   \thesaurus{
              (13.07.2;  
               08.16.6;  
               08.19.4;}  
\title{Gamma-ray sources as relics of recent supernovae in the nearby Gould
Belt}

\author{Isabelle A. Grenier}
\offprints{I. A. Grenier}
\institute{Universit\'e Paris VII and Service d'Astrophysique, CEA Saclay, 91191
Gif/Yvette, France\\ email: isabelle.grenier@cea.fr}

\date{Received October 6, 2000; accepted November 11, 2000}

\maketitle

\begin{abstract}
The nearby, 30 to 40 Myr old, starburst region of the Gould Belt
has formed numerous massive stars. Within its $\sim$300 pc radius,
it produces core-collapse supernovae at an enhanced rate which is
shown to be 75 to 95 Myr$^{ -1}$ kpc$^{-2}$, i.e. 3 to 5 times
higher than the local Galactic rate, over the past and future few
million years. A population of persistent, but unidentified,
$\gamma$-ray sources has been recently singled out at medium
latitudes above 100 MeV. Their distribution across the sky is
shown to be quite significantly and better correlated with the
tilted Gould Belt than with other Galactic structures. As many as
40 $\pm$ 5 sources are statistically associated with the Belt at
$|b|>5\degr$. It is therefore proposed that these sources are part
of the Belt and are relics of the Belt supernovae in the form of
million-year old pulsars. Their presence stresses how active the
local medium is, heated, enriched and shaped by multiple recent
explosions.

\end{abstract}

\section{Introduction}
Above 100 MeV, EGRET has discovered $\sim$70 persistent,
unidentified, sources at medium latitudes that clearly differ from
the $\sim$40 sources seen close to the Galactic plane
(\cite{3eg}): they are significantly softer, fainter, and have a
steeper logN-logS function than at low latitudes (\cite{gehrels}).
They have been tentatively associated with the local interstellar
medium and the Gould Belt (\cite{iagintegral}; \cite{gehrels};
\cite{iagsnowbird}), but their nature remains a mystery. The
likely candidates present in the Galactic disc, such as pulsars,
supernova (SN) remnants, and OB associations, cannot account for
so many sources off the plane. Interestingly, the starburst Gould
Belt disc is tilted at $\sim 20 \degr$ to the Milky Way and close
enough for weak sources to be detected. Proposing an origin of the
sources in the Belt however requires one a) to confront their
spatial distribution with that of the Belt and other likely
Galactic structures given strong observational biases; b) to
evaluate the number of likely $\gamma$-ray emitters, i.e. compact
stars, formed in the Belt in the recent past. Both aspects are
analyzed below before discussing the possible nature of the
sources.

\section{Supernova rate in the Gould Belt}
Our Galaxy produces on average 2.5$^{+0.8}_{-0.4}$ SNe per century
(\cite{tammann}), $ \sim $85\% of which arise from the core
collapse of a massive star. This value and the stellar
distribution in the Galaxy imply a local rate of 20 events
Myr$^{-1}$ kpc$^{-2}$, in reasonable agreement with the 29
progenitors Myr$^{-1}$ kpc$^{-2}$ found with masses $>$ 8
M$_{\odot}$ within 1 kpc from the Sun (\cite{tammann}). This
region, however, includes the starburst Gould Belt with its 300 pc
radius. Today, the Belt hosts 432 $ \pm $ 15 progenitors with
masses $>$ 8 M$_{\odot}$ (\cite{comeron}) and their maximum
lifetime will imply a crude minimum rate $>$ 40 collapses
Myr$^{-1}$ kpc$^{-2}$ in a few tens of Myr. Numerous explosions
have lately occurred from the first generations of massive stars
formed in the Belt. A recent-past rate can be inferred from the
current stellar content of the Belt as a function of mass, for
both short-lived and long-lived stars, given a few assumptions: 1)
a stellar initial mass spectrum $dN/dM= a \, M^{\Gamma-1}$ with a
range of indices $-2.0 \le \Gamma \le -1.1$ as measured in nearby
OB associations (\cite{scalo}; \cite{massey}); 2) lifetime
estimates as modelled for stars with solar metallicity ($Z=0.02$;
\cite{schaller}; \cite{meynet}) and interpolated in mass; 3) a
constant birth rate for simplicity; 4) a conservative mass
threshold for collapse of 8 M$_{\odot}$. Using this formalism,
star counts were integrated in various spectral bands and scaled
to the observed Belt star counts (\cite{comeron}) to eliminate the
unknown birth rate and amplitude \textit{a}.

The significant increase in birth rate obtained for stars $<$40
Myr confirms the Belt nuclear age and the validity of our simple
formalism. SN yields of 21.4 $ \pm $ 0.8, 24.1 $ \pm $ 0.9, and
27.2 $ \pm $ 1.0 per Myr were obtained for $\Gamma$ indices of
$-2.0$, $-1.5$, and $-1.1$, respectively, for a 40 Myr old Belt.
The quoted error results from the uncertainty in the observed star
counts. These estimates decrease by 15\% with Belt age from 50 to
30 Myr. Truncating the mass spectrum at 60 or 120 M$_{\odot}$, or
choosing 10 M$_{\odot}$ for the collapse threshold, has $<10\%$
impact on the results. Given the uncertainties for the Belt age
and $\Gamma$ index (particularly at large masses), we infer a
frequency of 20 to 27 SNe per Myr and a rate of 75 to 95
Myr$^{-1}$ kpc$^{-2}$ which is 3 to 5 times the local Galactic
one. This high rate is valid for the past few Myr. It is
consistent with the power of $2.3\times 10^{51}$ erg Myr$^{-1}$
kpc$^{-2}$ required to maintain the local cosmic-ray density
(\cite{blandford}) for a standard SN-to-cosmic-ray energy
conversion efficiency of a few percent (\cite{drury}). It is
consistent with the presence, within the Belt, of four 0.1-1 Myr
radio loops (\cite{berk}) and the Local Bubble. Thirteen radio
pulsars from the Princeton catalogue are found at high latitudes
with distances $<$1 kpc and age $<$2 Myr, but they are too few to
show a correlation with the inclined Belt or with the Galactic
plane. The narrow radio beams from many more may miss the Earth.

\begin{figure}
\resizebox{8.8cm}{!}{\includegraphics{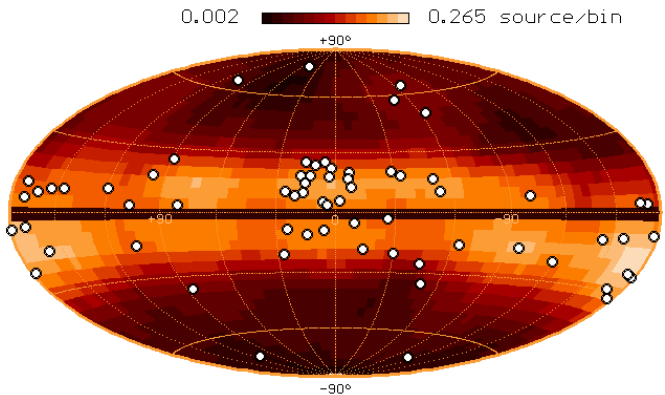}}
\caption{visibility map, in Galactic coordinates, from the best
fit ``Iso+Belt'' model, the Belt being traced by its young massive
stars. The colour codes the number of sources detectable per bin
in the EGRET survey. The 67 persistent, unidentified EGRET sources
at $|b|> 2.5\degr$ are marked as white dots which closely follow
the curved lane of Belt stars.} \label{map}
\end{figure}

\section{Gamma-ray sources in the Gould Belt}
The distribution on the sky of a subset of the unidentified EGRET
sources is strongly reminiscent of the tilted Gould Belt. The
subset includes all \textit{persistent} sources at $|b|>2.5\degr$
from the 3rd EGRET catalogue, i.e. those detected with a
significance $>4\sigma $ ($>5\sigma$ at $|b|<10\degr$) using the
cumulative data from April 1991 to October 1995. Sources listed as
likely artifacts were discarded. The 67 sources thus selected are
displayed in Fig. 1. 80\% of them show no time variability, the
others are only moderately variable (\cite{tompkins}). Strong
biases due to the instrument exposure and the bright interstellar
background does influence their apparent spatial distribution. To
assess their correlation with various structures, a
maximum-likelihood test was applied that takes these biases into
account (\cite{iagintegral}). An important degree of freedom is
the unknown luminosity function $dN/dL \propto$ L$^{\alpha}$ of
the parent sources. The flatter the function, the sharper the
longitude and latitude profiles of any Galactic population because
more sources remain visible to large distances. This effect being
quite strong, $\alpha$ was kept as a free parameter over a wide
range of values: $-2.5 <\alpha \le 0$. EGRET exposure maps above
100 MeV for the 4.5-year survey (\cite{3eg}), the observed
isotropic background of ($2.1 \pm 0.3)\times 10^{-5} \gamma$
cm$^{-2}$ s$^{-1}$ sr$^{-1}$, and the interstellar background
model (\cite{hunter}) of the EGRET survey were used to model the
source visibility in $5\degr$ by $5\degr$ bins across the sky.

To allow for possible extragalactic sources among the unidentified
ones, linear combinations of an isotropic and a Galactic component
were tested (and nicknamed as in Table 1). The choice of Galactic
distributions cover the various sites likely to display sources at
mid latitudes: 1) sources uniformly distributed in a spherical
Galactic halo, 20 kpc in radius; 2) sources uniformly distributed
in longitude in the ``local Galactic disc'' with \textit{any}
exponential or gaussian scale height; 3) sources spread in a
``thick Galaxy'' with a gaussian radial scale length of 9.3 kpc
and an exponential scale height of 0.4 kpc, typical of radio
pulsars (\cite{lyne}); 4) sources distributed in the interstellar
medium as mapped by the NH$=$N(HI+2H$_{2}$) column-densities
(\cite{iagintegral}); 5) sources in the Gould Belt as traced by
the column-densities of stars with spectral type $<$ B3, fitted
from the Yale Bright Star catalogue data according to the function
$N_{*}(l,b) \propto$
exp(-0.5*sin$^{2}[b-b_{*}(l)]$/sin$^{2}\delta$), with the
resulting width $\delta = 12.0\degr \pm 0.4\degr$ and median
latitude $b_{*}(l)$ of the star distribution displayed in Fig. 1.
Distributions 4) and 5) provide independent ways to trace the Belt
through its young stars as well as its clouds since the latter
dominate the interstellar maps at medium latitudes.

The reliability of the analysis was checked by applying it to the
67 active galactic nuclei discovered by EGRET and spread over the
sky. For them, an isotropic distribution indeed yields the best
fit and no Galactic component, however large in scale height, is
detected: the maximum log-likelihood value remains constant within
0.1 for all the models tested in Table 1. 67 $\pm$ 8 sources are
attributed to the isotropic component, 0 $\pm$ 6 to the local
Gal., thick Gal., NH, or Belt components. The 9 $\pm$ 16 sources
attributed to a halo component are not significant. The luminosity
index of the parent population is $\alpha = 1.7 \pm 0.3$.

\begin{table*}
\caption[]{max-likelihood results and their 1$\sigma $ errors for
selected models against the 67 persistent unidentified sources}
\begin{center}
\begin{tabular}{llllll}
\hline\noalign{\smallskip}
model & $P_{\alpha}[\textrm{model}|\textrm{ISO}]$& $N_{\textrm{iso}}$ & $N_{\textrm{ani}}$ & $N_{\textrm{ani}}$ & $\alpha$\\
& & $|b|>2.5\degr$ & $|b|>2.5\degr$ & $5\degr<|b|<30\degr$\\
\noalign{\smallskip}
\hline\noalign{\smallskip}
ISO=isotropic && 67 $\pm$ 8 & & & $>-0.5$\\
HAL=ISO+halo & 10$^{-4}$ & 41 $\pm$ 7 & 26 $\pm$ 7 & 14 $\pm$ 4 & $>-0.15$\\
GAL=ISO+thick Gal. & $2\times 10^{-9}$ & 0 $\pm$ 11 & 67 $\pm$ 11 & 39 $\pm$ 6 & $-1.90 \pm 0.07$\\
LOC=ISO+local Gal. & $7\times 10^{-9}$ & 0 $\pm$ 10 & 67 $\pm$ 10 & 38 $\pm$ 6 & $-1.90 \pm 0.06$\\
ISM=ISO+NH & $2\times 10^{-13}$ & 0 $\pm$ 6 & 67 $\pm$ 6 & 39 $\pm$ 4 & $-2.1 \pm 0.3$\\
BELT=ISO+Gould Belt & $2\times 10^{-14}$ & 12 $\pm$ 4 & 55 $\pm$ 4 & 47 $\pm$ 4 & $-2.2 \pm 0.3$\\
\noalign{\smallskip}
\hline
\end{tabular}
\end{center}
\end{table*}
\begin{figure}
\resizebox{8.5cm}{!}{\includegraphics{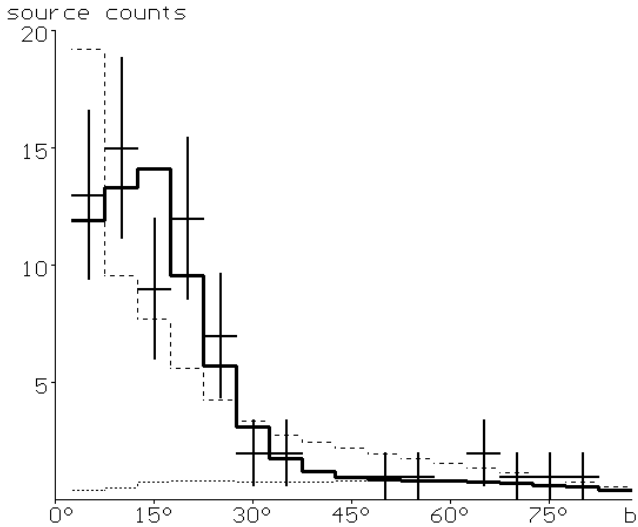}}
\caption{latitude profiles of the persistent unidentified EGRET
sources (crosses), of the best fit ``Iso+Belt'' model (thick
line), and of the isotropic contribution to this model (dotted
line). The ``Iso+Local Gal.'' model (dashed line) yields a
significantly poorer fit ($P_{\alpha}[\textrm{BELT}|\textrm{LOC}]=
\, 3\times 10^{-6}$) with systematic deviations from the data over
wide latitude bands.} \label{profile}
\end{figure}

Results obtained for the 67 unidentified sources are displayed in
Table 1. Modelled source counts in the isotropic and anisotropic
components, $N_{\textrm{iso}}$ and $N_{\textrm{ani}}$
respectively, were summed over the given latitude intervals. The
probability $P_{\alpha}[2|1]$ is determined from the
log-likelihood increase between two models, with $\alpha$ left
free. It is the chance probability that a random fluctuation from
model 1 yields as good a fit as model 2. The dramatic improvement
in the fit for any Galactic model over a pure isotropic population
shows that over $\sim$50 sources have a Galactic origin. The fit
improves very significantly from a spherical halo to a flatter
Galactic distribution, $P_{\alpha}[\textrm{GAL}|\textrm{HAL}]= \,
2\times 10^{-5}$, and even more so with the tilted geometry of the
Belt clouds or stars, $P_{\alpha}[\textrm{BELT}|\textrm{HAL}]= \,
2\times 10^{-10}$. The good quality of the BELT fit is illustrated
in Fig. 1 \& 2. The fact that the LOC and GAL models yield equally
good fits implies a large fraction of \textit{nearby} sources: no
contrast in longitude is detected that would require distances of
a few kpc to the sources. Observed source counts are indeed
equivalent in the centre and anticentre quadrants in Fig. 1. Not
only are the sources local, the significant likelihood increase
between the LOC and BELT fits, with a chance probability
$P_{\alpha}[\textrm{BELT}|\textrm{LOC}]= \, 3\times 10^{-6}$,
gives strong support to their origin in the inclined Belt system.
On the other hand, the sources at $|b|>5\degr$ were shown to be
distinctly fainter and softer than those at lower latitudes and a
subset of $\sim$20 were pointed out along the Belt
(\cite{gehrels}). Together these findings provide compelling
evidence that a distinct population of 20 to 40 EGRET sources
belong to the Gould Belt. Based on their spatial distribution
only, they could be as numerous as 40 $\pm$ 5 at $|b|>5\degr$. The
luminosity index of their parent population is $\alpha = 2.2 \pm
0.3$.

\section{Nature of the sources}
With luminosities $L_{\gamma}$ of 1 to $15\times 10^{26}$ W over
4$\pi$ sr at 500 pc (for E$^{-2}$ spectra), the sources cannot be
unresolved gas clumps irradiated by the local cosmic-ray flux: the
required mass of $\sim 10^{4}$ M$_{\odot}$ at 500 pc cannot have
escaped the radio and IR surveys, even when considering radio beam
dilution from unresolved clouds (\cite{iagintegral}). Nor can they
be slow ($<$20 km/s), old neutron stars, accreting gas from a
dense cloud: they are at least 10$^{2-3}$ times too rare
(\cite{iagintegral}) and the maximum Bondi-Hoyle accretion power
of $\sim 2\times 10^{25}$ W that results from the formation of a
surrounding HII region by the neutron star UV radiation, is 10
times too low (\cite{blaes}). The accretion power reached for fast
($>$200-400 km/s), highly magnetized (10$^{12}$ G) neutron stars
with long periods in the intercloud medium (10$^{-3}$ H
cm$^{-3}$), though increased by Kelvin-Helmholtz instabilities in
the shocked gas, is also orders of magnitude too low
(\cite{hardinglev}). Isolated accreting black holes with masses of
10 M$_{\odot}$ (\cite{colpi}) and 35 M$_{\odot}$ (\cite{dermer})
have been proposed, but they would be too rare in the Belt: for
mass progenitors $>$ 25 M$ \sun $ (\cite{timmes}), black holes are
3 to 9 times fewer than neutron stars for $\Gamma$ indices of
$-1.1$ \& $-2.0$, respectively. Moreover, the luminosity
$L_{\gamma}$/$L_{\textrm{X}}$ ratios $\gg 50$ observed assuming
statistically 1 or 2 faint ROSAT source in an EGRET error box, are
clearly at variance with p-p interactions in the accretion flow
(\cite{colpi}), unless non-thermal acceleration is advocated (in
micro-quasar jets?). These luminosity ratios are also at variance
with standard accreting binary systems. I found no spatial
coincidence with a WR star or the numerous O stars at mid latitude
despite their highly supersonic winds with kinetic powers of
10$^{28-29}$ W. Eight pulsars are known in $\gamma$ rays, 7 bright
young ones at kpc distances in the Galactic disc, and a faint,
older one inside the Gould Belt (Geminga). So, the present sample
is strongly biased to high luminosity and youth. The stability of
most of the Belt sources (\cite{tompkins}, \cite{gehrels}) is
consistent with a pulsar origin. Born with large velocities
(\cite{lynelorimer}), Galactic pulsars rapidly migrate away from
the plane to mid-latitudes. The outer gap model (\cite{yadi}) for
beamed emission predicts that 4-5 Galactic pulsars should be
detectable at $|b|>5\degr$ in contrast with the 40 $\pm$ 5 sources
associated with the Belt. Similarly, the wide-beam comptonized
polar-cap model (\cite{sturner}) predicts 1-2 Galactic pulsars
detectable at $|b|>10\degr$ as opposed to 25 $\pm$ 5 sources
linked to the Belt. These discrepancies cannot be resolved by
increasing the Galactic pulsar birth rate by more than 30\% for
fear of overproducing sources at low latitude (\cite{yadi}), nor
by using larger scale heights or velocities which are not
supported by the radio data.

Given the enhanced SN rate in the Belt and its inclined geometry,
I propose that the sources associated with the Belt be relics of
Belt supernovae in the form of few Myr old pulsars. Detecting 20
to 40 Belt collapsed stars as EGRET sources requires the product
of the beaming fraction $\Delta\Omega/4\pi$ and pulsar age be of
order 1-1.5, for instance $\Delta\Omega/4\pi = 0.5$ over 2 or 3
Myr. $\Delta\Omega/4\pi \sim 0.1$ is predicted for the main
polar-cap beam (\cite{thompson}), and values of 0.1-0.6 are
possible for the outer-gap fan beam depending on pulsar age
(\cite{romani}). Yet, one should bear in mind the extreme
closeness of the Belt objects. The $\gamma$-ray luminosity,
$L_{\gamma}$, scales with the spin-down power, $\dot{E}$, as
$L_{\gamma} \propto \dot{E}^{1/2}$ over 4 decades in $\dot{E}$
(\cite{thompson}). Extrapolating from the faint Geminga for only
half a decade, to $\dot{E} = 10^{27}$ W $\Longrightarrow
L_{\gamma} \sim 6\times 10^{25}$ W over 1 sr, suggests that a
pulsar 10 times as old as Geminga, i.e. 3-Myr old, remains easily
detectable by EGRET out to 500 pc. Furthermore, the recorded
$\gamma$-ray lightcurves show that 10 times fainter emission is
detected off the main peaks over large phase intervals. This side
emission from a 3 Myr old pulsar would remain detectable by EGRET
up to 350 pc, thus largely increasing the detection probability
$\Delta\Omega/4\pi$. Recent polar cap simulations indicate that
4-5 times as many off-beam sources as on-beam ones would be
detectable above 100 MeV at a given distance (\cite{harding}),
therefore up to 350 pc for side emission. In this case, the
population of Belt neutron stars born in the last 2-3 Myr may
account for the Belt $\gamma$-ray sources. It may also explain the
scarcity of bright on-beam-like sources off the Galactic plane.
The softness of side emission ($|\gamma|$ = 1.8 to 2.5) may also
explain the soft average spectral index $|\overline{\gamma}|$ =
2.25 $\pm$ 0.03 measured for the Belt sources as opposed to that
of $|\overline{\gamma}|$ = 1.74 $\pm$ 0.02 obtained the 5 on-beam
pulsars in the Galactic plane. Preliminary simulations show that
the Belt spatial signature is preserved over at least 2 Myr
despite rapid migration. So, given our present understanding of
pulsar $\gamma$-ray emission, the hypothesis that the Gould Belt
sources be powered by pulsars, mostly off-beam pulsars, is quite
plausible. If true, the Belt pulsars would considerably broaden
our understanding of photon-particle cascades inside their
magnetospheres to older and lower-luminosity objects at various
aspect angles. The radio beam being apparently much narrower than
the $\gamma$-ray beam, at least in one dimension, one would expect
a majority of radio-silent pulsars among the Belt sources. The
spectral criterion $|\gamma|<$ 2 often adopted to search for
$\gamma$-ray pulsars may not be valid for nearby objects. The next
generation telescope, GLAST, to be launched in 2005, will be able
to detect their periodicity in the $\gamma$-ray signal, if any. As
supernova relics, the Belt sources would bring useful constraints
on the initial star mass spectrum at large masses and on the
abundance of explosive nucleosynthesis products in supernova
remnants.

\begin{acknowledgements}
      With sincere thanks to C. Dermer, N. Gehrels, A. Harding, P. Michelson, J.
Paul and D. Thompson.
\end{acknowledgements}


\begin{thebibliography}{}

\bibitem[Berkhuijsen 1973]{berk} Berkhuijsen E.M., 1973, A\&A 24, 143

\bibitem[Blaes et al. 1995]{blaes} Blaes O., Warren O., Madau P., 1995, ApJ 454, 370

\bibitem[Blandford \& Ostriker 1980]{blandford} Blandford R.D., Ostriker J.P., 1980, ApJ 237, 793

\bibitem[Colpi et al. 1986]{colpi} Colpi M., Maraschi L., Treves A., 1986, ApJ, 311, 150

\bibitem[Comeron et al. 1994]{comeron} Comeron F., Torra J., Gomez A.E., 1994, A\&A 286, 789

\bibitem[Dermer 1997]{dermer} Dermer C.D., 1997, Proc. 4$^{th}$ Compton Symp., AIP 410, 1275

\bibitem[Drury et al. 1994]{drury} Drury L.O'c., Aharonian F.A., V\"olk H.J., 1994, A\&A 287, 959

\bibitem[Gehrels et al. 2000]{gehrels}Gehrels N., Macomb D.J., Bertsch D.L., Thompson D.J., Hartman R.C., 2000, Nature 404, 363

\bibitem[Grenier 1997]{iagintegral} Grenier I.A., 1997, Proc. 2$^{nd}$ Integral Workshop, The Transparent Universe, ESA-SP382 187

\bibitem[Grenier 2000]{iagsnowbird} Grenier I.A., 2000, Proc. 6$^{th}$ Toward a major atmospheric Cherenkov detector Symp., AIP 515,
261

\bibitem[Harding \& Zhang 2001]{harding} Harding A.K., Zhang B., 2001, ApJ Letters, in press

\bibitem[Harding \& Leventhal 1992]{hardinglev} Harding A.K., Leventhal M., 1992, Nature 357, 388

\bibitem[Hartman et al. 1999]{3eg} Hartman R.C., Bertsch D.L., Bloom S.D., et al., 1999, ApJS 123, 79

\bibitem[Hunter et al. 1997]{hunter} Hunter S.D., Bertsch D.L., Catelli J.R., et al., 1997, ApJ 481, 20

\bibitem[Lyne et al. 1985]{lyne} Lyne A.G., Manchester R.N., Taylor J.H., 1985, MNRAS 213, 613

\bibitem[Lyne \& Lorimer 1994]{lynelorimer} Lyne A.G., Lorimer D.R., 1994, Nature 369, 127

\bibitem[Massey et al. 1995]{massey} Massey P., Johnson K.E., DeGioia-Eastwood K., 1995, ApJ 454, 151

\bibitem[Meynet et al. 1994]{meynet} Meynet G., Maeder A., Schaller G., Schaerer D., Charbonnel C., 1994, A\&AS 103, 97

\bibitem[P\"oppel 1997]{poppel} P\"oppel W.G.L., 1997, Fund. of Cosmic Physics 18, 1

\bibitem[Romani 1996]{romani} Romani R.W., 1996, ApJ 470, 469

\bibitem[Scalo 1986]{scalo} Scalo J.M., 1986, Fund. of Cosmic Physics 11, 1

\bibitem[Schaller et al. 1992]{schaller} Schaller G., Schaerer D., Meynet G., Maeder A., A\&AS 96, 269

\bibitem[Sturner \& Dermer 1996]{sturner} Sturner S.J., Dermer C.D., 1996, ApJ 461, 872

\bibitem[Tammann et al. 1994]{tammann} Tammann G.A., L\"offler W., Schr\"oder A., 1994, ApJ 92, 487

\bibitem[Thompson et al. 1997]{thompson} Thompson D.J., Harding A. K., Hermsen W., Ulmer M. P., 1997, Proc. 4$^{th}$ Compton Symp., AIP 410, 39

\bibitem[Timmes et al. 1996]{timmes} Timmes F.X., Woosley S.E., Weaver T.A., 1996, ApJ 457,
834

\bibitem[Tompkins 1999]{tompkins} Tompkins B., 1999, PhD thesis, Stanford University

\bibitem[Yadigaroglu \& Romani 1997]{yadi} Yadigaroglu I.A., Romani R.W., 1997, ApJ 476, 347

\end{thebibliography}
\end{document}